\documentclass{article}
\usepackage{graphicx}

\newcommand{\beq}{\begin{equation}}
\newcommand{\eeq}{\end{equation}}
\newcommand{\beqn}{\begin{eqnarray}}
\newcommand{\eeqn}{\end{eqnarray}}
\newcommand{\nbeqn}{\begin{eqnarray*}}
\newcommand{\neeqn}{\end{eqnarray*}}
\newcommand{\bfig}{\begin{figure}}
\newcommand{\efig}{\end{figure}}
\newcommand{\btab}{\begin{table}}
\newcommand{\etab}{\end{table}}
\newcommand{\benu}{\begin{enumerate}}
\newcommand{\eenu}{\end{enumerate}}
\newcommand{\bite}{\begin{itemize}}
\newcommand{\eite}{\end{itemize}}
\newcommand{\bdis}{\begin{displaymath}}
\newcommand{\edis}{\end{displaymath}}
\newcommand{\bary}{\begin{array}}
\newcommand{\eary}{\end{array}}

\newcommand{\ud}{\mathrm{d}}

\begin{document}
\begin{titlepage}
%\begin{flushright}
%Draft \\
%\today
%\end{flushright}
\vspace*{2in}
\begin{center}
\Large{\bf A new  evaluation of the Baldin sum rule}\\ [30pt] 
\large {D.Babusci, G.Giordano, G.Matone} \\[15pt]
{\it Istituto Nazionale di Fisica Nucleare - Laboratori Nazionali 
di Frascati, P.O. Box 13, I-00044 Frascati (Rome), Italy} \\ [80pt]
\end{center}
\begin{center}
{\bf Abstract}
\end{center}
\vspace*{0.5cm}
The Baldin sum rule for the nucleon has been recalculated at the 
light of the most recent photoabsorption cross section measurements.
The proton value $(\alpha + \beta)_p \,=\,13.69\,\pm\,0.14$ is smaller
but consistent with the one usually quoted in literature. However,
the value for the neutron $(\alpha + \beta)_n \,=\,14.40\,\pm\,0.66$
turns out to be three standard deviations away from the previously
calculated one. \\[13pt]
PACS number(s): 13.60.Hb, 14.20.Dh \\[150pt] 
\begin{center}
Submitted to Physical Review C Rapid Communication
\end{center}
\end{titlepage}

The optical theorem applied to the forward Compton amplitude, 
together with the low energy theorem, leads to the once-subtracted 
dispersion relation worldwide known as the Baldin sum rule 
\cite{BaLa}. This equation establishes a firm connection between 
the integral of the $\nu^2$-weighted nucleon unpolarized 
photoabsorption cross section and the sum of the electric ($\alpha$) 
and magnetic ($\beta$) polarizabilities of the nucleon target:
\beq\label{eqSR}
(\alpha\,+\,\beta)_N\,=\,\frac{1}{2 \pi^2}\,\int_{\nu_0}^\infty\,
\ud \nu\,\frac{\sigma(\gamma N \;\to\;X)}{\nu^2}
\eeq
where $\nu_0$ is the pion photoproduction threshold.
Since the integral on the right hand side can be numerically 
evaluated on the basis of the photoabsorption cross section 
data, eq.(\ref{eqSR}) leads to an unavoidable bound, that, 
as such, is routinely used to constrain the values of the 
polarizabilities extracted from the low energy Compton 
scattering data.

The numerical value quoted in literature for the proton
\footnote{hereafter the polarizability values are expressed in 
units of 10$^{-4}$ fm$^3$}
\beq\label{eqDG}
(\,\alpha\,+\,\beta\,)_p\,=\,14.2\,\pm\,0.3\;,
\eeq
was calculated over 25 years ago by Damashek and Gilman \cite{DaGi}.
They used the experimental data available at that time and 
postulated a reasonable theoretical "ansatz" for the extrapolation 
at infinite energy whose uncertainty is what fully determines the error
bar quoted in eq.(\ref{eqDG}), without taking into account any other 
source of errors.

As for the neutron, the first, and still unique, complete calculation 
of the sum rule was made in 1979 by the authors of ref. \cite{LvPe}. 
In this calculation the integration domain is broken down into a 
resonance ($\nu \le$ 1.5 GeV) and an asymptotic ($\nu >$ 1.5 GeV) 
region. In the first region, they used a multipole analysis of the 
single-pion photoproduction data and assumed that the two pion 
contributions were dominated by the leading $\Delta$ and 
$\rho$-meson photoproduction channels. By using the parametrization 
given in ref. \cite{ArmP} for the asymptotic regime, they finally 
obtained:
\beq\label{eqLP}
(\,\alpha\,+\,\beta\,)_n\,=\,15.8\,\pm\,0.5\;.
\eeq

Since today the status of the experimental data is much better defined 
than it was 20 years ago, it is now time to revisit the analysis for both 
the values of eq.(\ref{eqDG}) and (\ref{eqLP}). Let us discuss the two 
cases separately.
\vspace*{0.3cm}

\section*{The Proton}

The integration domain has been divided into the following four energy 
regions:
\bite
\item the threshold region ${\cal A}^{\,(p)}$ : 
$\nu \in [\nu_0\;,\;0.2)$ GeV
\item the resonance region ${\cal B}^{\,(p)}$ : 
$\nu \in [0.2\;,\;2.0)$ GeV
\item the high-energy region ${\cal C}^{\,(p)}$ : 
$\nu \in [2.0\;,\;183.0)$ GeV
\item the asymptotic region ${\cal D}^{\,(p)}$ : 
$\nu \in [183.0\;,\;\infty)$ GeV
\eite

In the threshold region the total cross section has been calculated 
by a numerical integration of the $\pi^0 p$ and $\pi^+ n$ contributions 
given by the SAID program (solution SP97K) \cite{SAID}. The finite 
spacing between the points (1 MeV) generates an uncertainty in the 
evaluation of the subtended area which reflects itself in the error 
quoted in table I for ($\alpha + \beta$) in this region.

In the resonance region we have used the old values for the total cross
section measured at Daresbury \cite{ArmP} and the new data recently
obtained at Mainz \cite{McCo} in the interval 
$\nu \in (204\,,\,789)$ MeV. All these data (for a total of 138 
points) have been fitted using a minimizing function written as a 
sum of the six prominent Breit$-$Wigner resonances P$_{33}$(1232), 
P$_{11}$(1440), D$_{13}$(1520), S$_{11}$(1535), F$_{15}$(1680), 
F$_{37}$(1950) and a smooth background parametrized as follows 
\cite{ArmP}: 
\beq\label{eqBGres}
\sigma_{\mathrm{B}}\,=\,\sum_{k = -2}^{2}\,C_k\,(W - W_0)^k 
\eeq
where $W = M_p \sqrt{1 + 2 \nu/M_p}$ is the center of mass energy 
and $W_0 = W(\nu = \nu_0)$. However our major interest has not been
focused on the extraction of the resonance parameters but only on 
the determination of the most faithful mathematical description 
of the data. As a consequence of the $\nu^2$-weight in eq.(\ref{eqSR}), 
this description must be particularly accurate in the low-energy 
region. Therefore, instead of using the parametrization of ref. 
\cite{Bian}, we have adopted eq.(\ref{eqBGres}) as a description 
of the non-resonant pionic background. This choice produces a lower 
reduced $\chi^2_{df}$ and a more accurate description of the 
behaviour of the data on the rise of the $\Delta$-resonance. 
Only the statistical errors have been considered.

The complete collection of the data in the resonance region together 
with our fitting curve are shown in fig.(1).
\bfig[!hb]\label{fig:pro}
\vspace*{-1.5cm}
\centering
\includegraphics[width=5.0in]{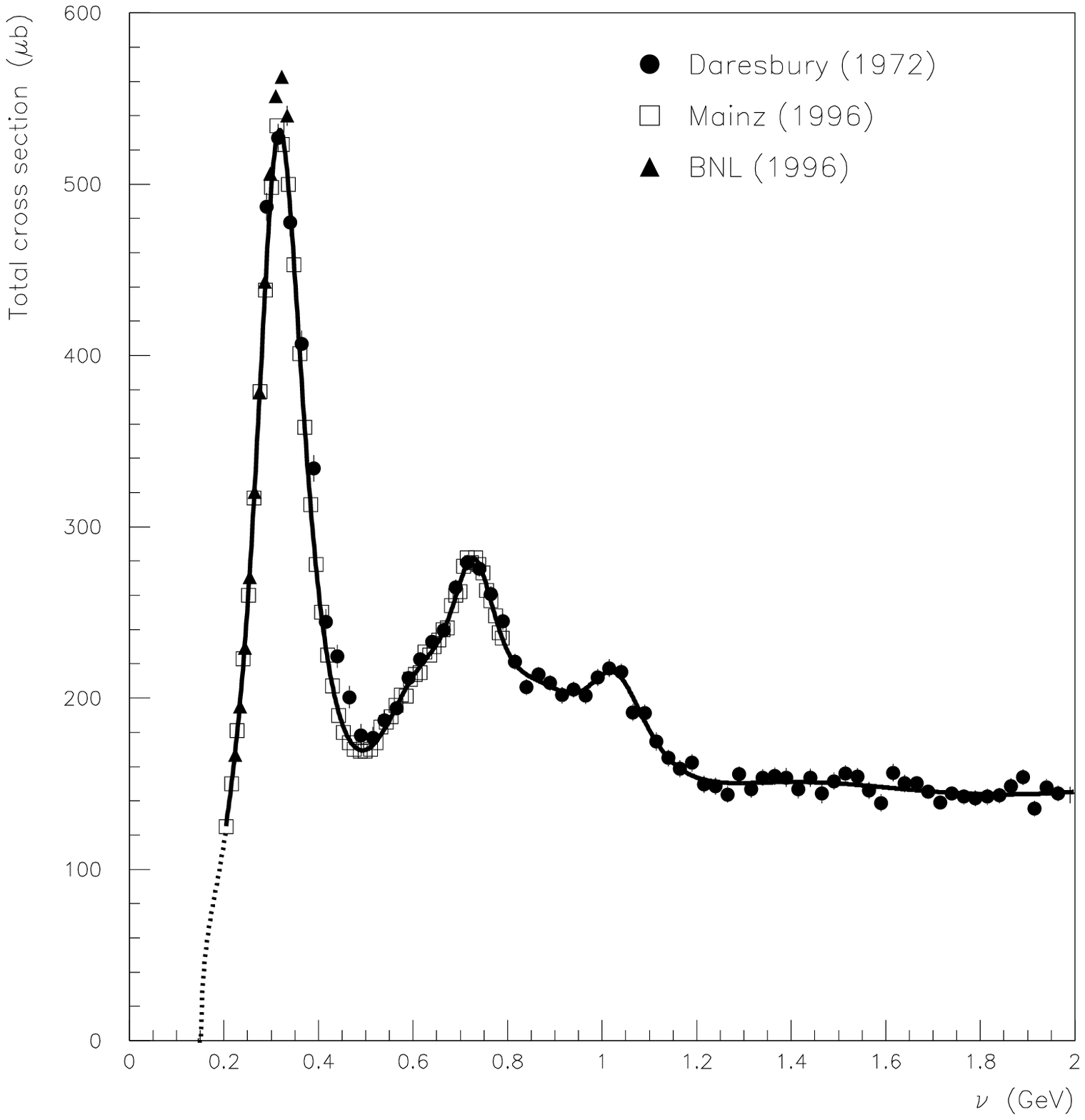}
\vspace*{-0.5cm}
\caption{Photoabsorption cross section for the proton as a function 
of the energy of the incoming photon. The dotted line is the result 
obtained from the SAID program in the threshold region.}
\efig

Since below about 400 MeV, the absorption cross section is completely 
dominated by single pion photoproduction, an independent measurement 
of the total cross section can also be deduced from the multipole 
analysis of $\gamma N \to \pi N$. However, pion production experiments 
between 1970 and 1980 display an unusual dichotomy near the peak of 
$\Delta$-resonance. For photon energies either below 280 MeV or 
above 360 MeV, $\pi^+$ and $\pi^0$ data taken at Bonn \cite{Bonn}, 
Tokyo \cite{Toky} and Lund \cite{Lund} are quite in good agreement.
Instead, within this energy range the Tokyo $\pi^+$ data and the 
Lund $\pi^0$ data are consistently higher than their Bonn 
counterparts. Since the recent Mainz absorption measurements 
\cite{McCo} are in good agreement with integrations of the Bonn 
$\pi^+$ and $\pi^0$ cross sections \cite{Pfei}, the Tokyo and
Lund data have fallen into general disfavor. However, very recent 
$\pi^+$ and $\pi^0$ cross sections measured by the LEGS collaboration
at BNL \cite{Sand} in the interval $\nu \in (210\,,\,333)$ MeV
are in fact in good agreement with the Tokyo and Lund data sets. 
Evidently the dichotomy at the $\Delta$-resonance still persists. 

To examine the consequences of the higher Tokyo/Lund/BNL cross sections 
we have repeated the fit in the resonance region, using the  total 
cross sections from the multipole analysis of the BNL data 
in place of the Daresbury and Mainz data below 340 MeV (in the 
following we shall refer to this as fit II). This fit departes from
the one displayed in fig.(1) only at the top of the 
$\Delta$-resonance where the total cross section turns out to be 
approximately 6 \% higher. The $\chi^2_{df}$ is slightly worse 
but the parameters of all the resonances involved are well 
reproduced within the errors.

According to ref. \cite{ArmP} in the region between 2 and 3 GeV, 
the cross section can be parametrized in the following way:
\beq\label{eqBGpio}
\sigma(\gamma p \;\to\;X)\,=\,a_1\,+\,\frac{a_2}{\sqrt{\nu}}
\qquad \mbox{with} \qquad
\bary{l}
a_{1} \,=\, 91.0\,\pm\,5.6\;\;\mu \mathrm{b} \\ 
a_{2} \,=\, 71.4\,\pm\,9.6\;\;\mu \mathrm{b} \cdot \mbox{GeV}^{1/2}
\eary
\eeq
An accurate fit of all the data avalaible in the remaining part of the 
region ${\cal C}$ can be found in ref. \cite{PDG1} where the following 
parametrization is used ($\nu$ in GeV):
\beq\label{eqrB}
\sigma(\gamma p \;\to\;X)\,=\,A\,+\,B\,\ln^2 \nu \,+\,C\,\ln \nu  
\qquad \mbox{with} \qquad
\bary{l}
 A\,=\,147\,\pm\,1\;\;\mu \mathrm{b} \\
 B\,=\,2.2\,\pm\,0.1\;\;\mu \mathrm{b} \\
 C\,=\,-17.0\,\pm\,0.7\;\;\mu \mathrm{b}
\eary
\eeq

This parametrization has been assumed valid also in the asymptotic 
region ${\cal D}$ and its result compared to the one given by the 
model of Donnachie and Landshoff where, for $\nu \ge$ 12 GeV, the total 
cross section is parametrized in this other way \cite{PDG2} ($s = W^2$ 
in GeV$^2$):
\beq\label{eqrDL}
\sigma(\gamma p \;\to\;X)\,=\,X\,s^\varepsilon \,+\,Y\,s^\eta 
\qquad \mbox{with} \qquad
\bary{ll}
 X = 71\,\pm\,18\;\;\mu \mathrm{b} & Y \,=\,120\,\pm\,40\;\;\mu \mathrm{b} \\
 \varepsilon\,=\,0.075\,\pm\,0.030 &  \eta\,=\,-\,0.46\,\pm\,0.25
\eary
\eeq

The contributions to $(\alpha + \beta)$ coming from the four regions 
defined above are reported in table I, where for the asymptotic region
the two values are the results obtained from eq.(\ref{eqrB}) (upper) 
and eq.(\ref{eqrDL}) (lower), respectively. By summing up these four 
contributions one has:
\beq\label{eqPro}
(\alpha\,+\,\beta)_p \,=\, 13.69 \,\pm\,0.14
\eeq
The use of the fit II in the resonance region would increase this value 
up to 13.76, well within the quoted error in eq.(\ref{eqPro}). Therefore 
the debate on the value of the total cross section at the top of the 
$\Delta$-resonance does not seem to be relevant for our present purpose.
\btab[!ht]
\begin{center}
Table I
\end{center}
\begin{center}
\begin{tabular}{||c||c|c|c|c||c||}
\hline \hline
 & & & & & \\ [2pt]
 Energy Region  & ${\cal A}^{\,(p)}$ & ${\cal B}^{\,(p)}$ &
 ${\cal C}^{\,(p)}$ & ${\cal D}^{\,(p)}$ & Total \\
 & & & & & \\ [2pt]
\hline \hline
 & & & & & \\ [2pt]
 & & & & $(7.0 \,\pm\, 0.3) \cdot 10^{-3} \quad$ & \\
$(\alpha + \beta)_p$ & $1.25 \,\pm\, 0.02$ & $11.71 \,\pm\, 0.13$ 
& $0.72 \,\pm\, 0.03$ & & $13.69 \,\pm\, 0.14$ \\ 
& & & & $(6.8 \,\pm\, 2.1) \cdot 10^{-3} \quad$ & \\
& & & & & \\ [2pt]
\hline \hline
\end{tabular}
\end{center}
\etab
\vspace*{0.3cm}

\section*{The Neutron}

The neutron case can be calculated by assuming that the  
photoabsorption cross section on the free neutron can be simply 
obtained by the ``difference'' between the deuteron and proton data.
The way to perform this difference is not firmly established and
can drive to evident inconsistencies. As an example, in the region 
of the $\Delta$-resonance the sum of the one-pion photoproduction 
cross sections \cite{SAID} alone is about 150 $\mu$b larger than the 
total absorption cross section published in ref. \cite{ArmD}. Since 
the photoabsorption cross section on the deuteron measured at Daresbury 
has been recently confirmed by the Mainz data \cite{McCo}, the 
discrepancy has to arise from the procedure followed to extract the 
neutron cross section from the deuteron data. This implies that further 
assumptions will be necessary with the consequence that the resulting 
value for $(\alpha + \beta)_n$ is much more procedure-dependent than 
that for $(\alpha + \beta)_p$.

Also in the deuteron case the energy range is divided in the four 
following regions:
\bite
\item the threshold region ${\cal A}^{\,(n)}$ : 
$\nu \in [\nu_0\;,\;0.2)$ GeV
\item the resonance region ${\cal B}^{\,(n)}$ : 
$\nu \in [0.2\;,\;2.0)$ GeV
\item the high-energy region ${\cal C}^{\,(n)}$ : 
$\nu \in [2.0\;,\;18.0)$ GeV
\item the asymptotic region ${\cal D}^{\,(p)}$ : 
$\nu \in [18.0\;,\;\infty)$ GeV
\eite
Similarly to the proton case, the total photoabsorption cross 
section in the threshold region results from the sum of the 
$\pi^- p$ and $\pi^0 n$ channels as given by the SAID program.

In the resonance region the avalaible data \cite{McCo,ArmD} for the
deuteron target have been fitted using the same procedure followed 
in ref. \cite{ArmD}. The minimizing function is the same as that for 
the proton with a non-resonant pionic background constrained to be 
twice the one found for the proton. Furthermore, following ref. 
\cite{Bian} we have added a deuteron photodisintegration background 
which gives a non-negligible contribution mainly to the $\Delta$-region 
\cite{BiMu}. The result of this fit together with the experimental data 
are shown in fig.2.
\bfig[!ht]\label{fig:deu}
\vspace*{-1.5cm}
\centering
\includegraphics[width=5.0in]{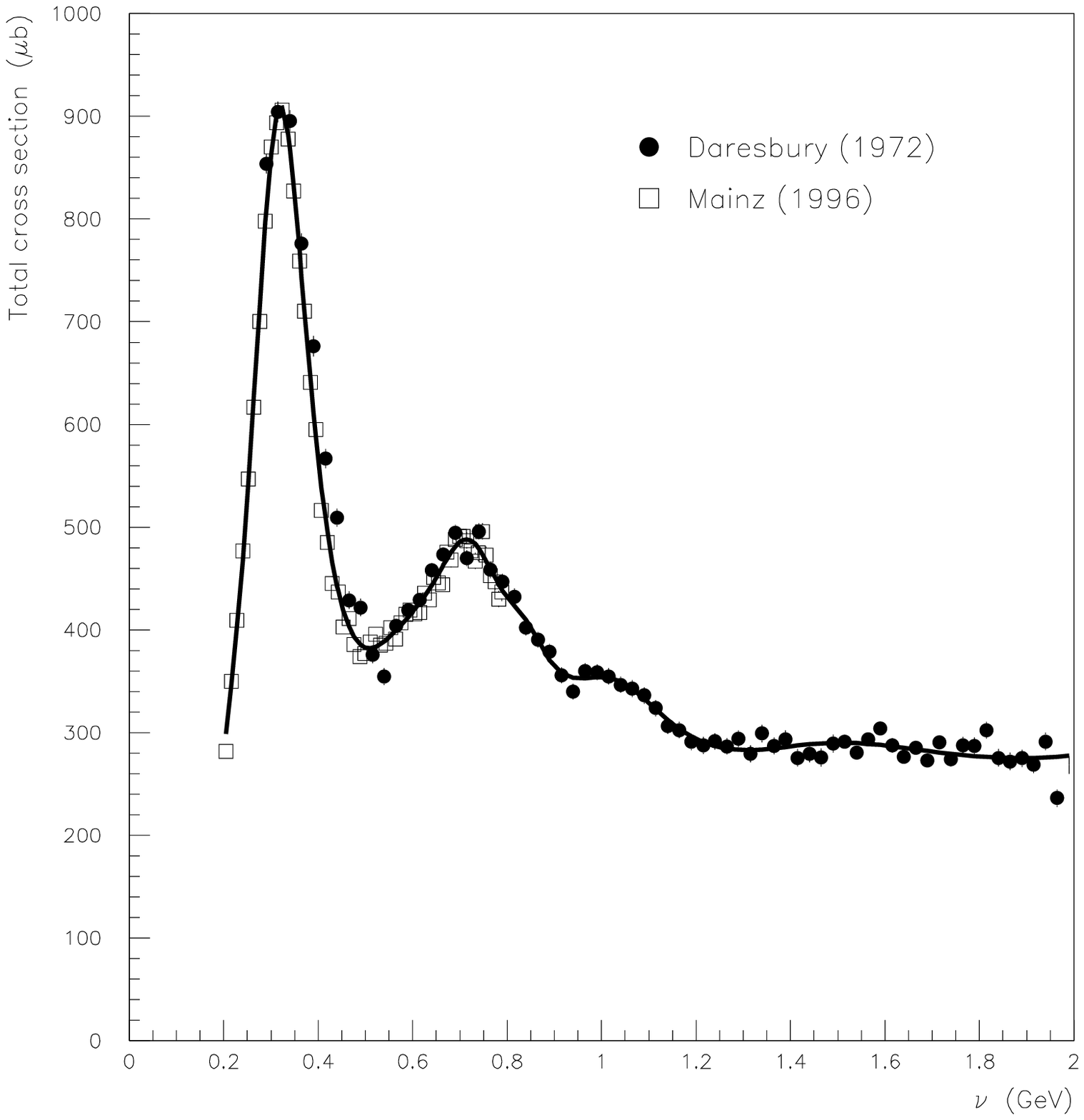}
\vspace*{-0.5cm}
\caption{As fig.1 in the deuteron case.}
\efig

The neutron cross section has been derived under the assumptions 
that the positions $W$ and widths $\Gamma$ of the resonances are 
the same for both the proton and the neutron and the coupling 
constants are related by:
\bdis
I_r^n\,=\,\frac{1}{\Gamma_r^p}\,\big(\,I_r^D\,\Gamma_r^D\,-\,
I_r^p\,\Gamma_r^p\,\big)\,.
\edis
The validity of these assumptions are discussed at length in ref. 
\cite{Bian}.

In the high-energy region ${\cal C}$ the photoabsorption cross section 
on deuteron can be parametrized by the expression of eq.(\ref{eqrB}) 
where \cite{PDG1}:
\bdis
A_D\,=\,300\,\pm\,5\;\;\mu \mathrm{b}\;, \qquad 
B_D\,=\,9.5\,\pm\,2.0\;\;\mu \mathrm{b}\;, \qquad 
C_D\,=\,-\,57\,\pm\,7\;\;\mu \mathrm{b}\;.
\edis
According to ref. \cite{ArmD} for $\nu \in (2\,,\,4)$ GeV it turns 
out that:
\beq\label{eqAD}
\sigma_n\,\simeq\,1.015\,\sigma_D\,-\,\sigma_p 
\eeq
Therefore, by assuming that this relationship can be extended to 
the whole region ${\cal C}$, for the neutron one has:
\bdis
A_n\,=\,157.5\,\pm\,5.2\;\;\mu \mathrm{b}\;, \qquad 
B_n\,=\,7.4\,\pm\,2.0\;\;\mu \mathrm{b}\;, \qquad
C_n\,=\,-\,40.9\,\pm\,7.1\;\;\mu \mathrm{b}\;.
\edis

Finally, in complete analogy with the proton case, we have assumed 
that the parametrization in the region ${\cal C}$ can be successfully 
extended to the asymptotic region ${\cal D}$.

The contributions to $(\alpha + \beta)$ coming from the four 
regions defined above are reported in table II and their sum is: 
\beq\label{eqNeu}
(\alpha\,+\,\beta)_n \,=\, 14.40 \,\pm\,0.66
\eeq
\btab[!ht]
\begin{center}
\vspace*{1.0cm}
Table II
\end{center}
\begin{center}
\begin{tabular}{||c||c|c|c|c||c||}
\hline \hline
 & & & & & \\ [2pt]
 Energy Region  & ${\cal A}^{\,(n)}$ & ${\cal B}^{\,(n)}$ &
 ${\cal C}^{\,(n)}$ & ${\cal D}^{\,(n)}$ & Total \\
 & & & & & \\ [2pt]
\hline \hline
 & & & & & \\ [2pt]
$(\alpha + \beta)_n$ & $1.86 \,\pm\, 0.02$ & $11.95 \,\pm\, 0.66$ 
& $0.52 \,\pm\, 0.05$ &  $0.07 \,\pm\, 0.02$ & $14.40 \,\pm\, 0.66$ \\
& & & & & \\ [2pt]
\hline \hline
\end{tabular}
\end{center}
\etab
\vspace*{0.3cm}

\section*{Summary and conclusions}

The present reevaluation of the sum rule (\ref{eqSR}) leads to the 
following conclusions:
\benu
\item[1) ] The new values of eqs.(\ref{eqPro},\ref{eqNeu}) are both
smaller than the corresponding values of eqs.(\ref{eqDG},\ref{eqLP})
but, within errors, they are still consistent with each other. The 
lower value for the proton could be due to the set of data used in 
the analysis of ref. \cite{DaGi} that consistently exceed the Daresbury 
data in the region of the P$_{11}$ and D$_{13}$ resonances.
\item[2) ] The proton and neutron values are much closer now than in  
the previous analysis. The present separation between these two values 
is within errors, whereas, before, the same separation was twice 
the sum of the quoted errors. This is consistent with the old 
claim reported in ref. \cite{MaPr} that no isotopic effect has to 
be expected for the quantity $\alpha + \beta$.
\item[3) ] Chiral perturbation theory at $O(q^4)$ gives the following
prediction for the sum rule \cite{Bern}: 
\bdis
(\alpha\,+\,\beta)_p\,=\,14.0\,\pm\,4.1\;,\qquad \qquad
(\alpha\,+\,\beta)_n\,=\,21.2\,\pm\,3.9
\edis
that is consistent with eq.(\ref{eqPro}) for the proton but appears to 
be ruled out by both the present and old analysis in the neutron case.
Instead, our combined proton/neutron result is much more in line with 
the $O(q^3)$ prediction which reads \cite{Bern}:
\bdis
(\alpha\,+\,\beta)_p\,=\,(\alpha\,+\,\beta)_n\,=\,13.3\;.
\edis
As a matter of fact, this value is well consistent with our proton 
value and is less than two standard deviations away from the neutron 
value.
\eenu
\vspace*{0.3cm}

\section*{Acknowledgements}

We would like to express our gratitude to N. Bianchi and  V. Muccifora 
for the information and the many useful discussions we had together on 
the whole subject. We also want to thank A. M. Sandorfi for providing
us with the total cross section for the proton obtained from the LEGS 
data.

\end{document}